\documentclass[letter]{aa}     
\usepackage{graphicx}
\usepackage{cuted}
\usepackage{capt-of}
\usepackage{txfonts}
\usepackage{lipsum}
\usepackage{subcaption}         
\usepackage{lscape}             
\usepackage{placeins}           
\newcommand{\kms}{{\mathrm{km~s^{-1}}}}

\begin{document}

   \title{Asteroseismology of blue straggler stars in $\omega$ Centauri}
   \subtitle{Stellar collisions as a possible formation scenario}

   \author{J. Daszyńska-Daszkiewicz\inst{1}\corrauth{jadwiga.daszynska-daszkiewicz@uwr.edu.pl}
        \and P. Góra\inst{1}
        \and P. Walczak\inst{1}
        \and A. Miszuda\inst{1}
        \and W. Szewczuk\inst{1}
        \and M. Heller\inst{2,3}
        \and F.R.N.Schneider\inst{3,2}
        }

 \institute{University of Wrocław, Faculty of Physics and Astronomy, Astronomical Institute, Kopernika 11, PL-51-622 Wrocław, Poland
 \and Zentrum für Astronomie der Universität Heidelberg, Astronomisches Rechen-Institut, Mönchhofstr. 12-14, Heidelberg, Germany
   \and Heidelberger Institut für Theoretische Studien, Schloss-Wolfsbrunnenweg 35, 69118 Heidelberg, Germany}

   \date{Received ..., Accepted}

  \abstract
   {Blue stragglers are intriguing stars that appear more massive and younger than expected from single-star evolution, yet their evolutionary histories remain elusive because of the multiple possible formation channels. A promising and poorly explored avenue for addressing this problem is asteroseismology, which  provides the most stringent constraints on internal structure and evolution.}
   {We investigate five SX Phoenicis stars from the core of the globular cluster $\omega$ Centauri that pulsate in two radial modes and occupy the blue-straggler region of the colour–magnitude diagram. Seismic models are constructed assuming a collisional origin and compared with our previous seismic studies based on single-star evolution. For comparison, we also explore a binary-evolution scenario involving mass transfer for one of the stars.} 
   {The evolution of the parent stars is computed with MESA. The collision is then modelled with the MMAMS code, which reconstructs the merger remnant using an entropic variable. The subsequent evolution is followed with MESA up to the current age of the cluster. Finally, we construct seismic models reproducing the observed frequencies of the two radial modes.}
   {We present the first seismic modelling of SX Phe stars (blue stragglers) based on the collision-formation scenario. Convective-core overshooting is essential for rejuvenation and for obtaining seismic models in the main-sequence phase of evolution, but it is not always sufficient. Whether a merger remnant undergoes rejuvenation depends in a complex way on the masses of the parent stars, metallicity, and collision time. We identify three evolutionary behaviours: no, delayed, and immediate rejuvenation. The collisional seismic  models imply younger ages and higher masses than the seismic models based on single-star evolution. }
   {}


   \maketitle
   \nolinenumbers

\section{Introduction}
Blue straggler stars (BSSs) are among the most challenging astrophysical objects for the theory of stellar formation and evolution.
This is because BSSs are more massive and much younger than the population they belong to and in which they formed \citep{Sandage1953}.
These facts contradict the single-star evolution, and blue stragglers appear to lag behind in their evolution.
BSSs can be easily identified in stellar clusters as their position on the colour-magnitude diagram (CMD) is well defined, i.e.,
they are bluer and brighter than stars at the main sequence (MS) turn-off \citep{Sandage1953}.

BSSs can be formed in more than one way, i.e., via mass transfer (MT) in binary systems 
\citep[e.g.,][]{McCrea1964,Ivanova2015}, coalescence in primordial binaries \citep[e.g.,][]{Nelson2001,Chen2008} or direct collisions of stars in dense cluster regions \citep[e.g.,][]{Benz1987,Sills2002}. Although the proposed formation scenarios are physically plausible and justified, distinguishing  between them using available observables remains a major challenge and does not usually lead to a unique solution.
For example, \citet{Ferraro2009} identified two sequences of BSSs on the CMD 
in the globular cluster M\,30, suggesting the existence of two distinct sub-populations. They proposed 
that the bluer sequence arose from direct collisions and 
the redder one arose from mass transfer and mergers in close binaries. 
This interpretation was further supported by \citet{PortegiesZwart2019}, who modelled stellar mergers formed either through direct collisions or during an unstable phase of mass transfer in binary systems.
However, the most recent  studies by  \citet{Kumawat2026} of 56 Milky Way globular clusters (including M\,30) 
showed no statistical evidence for double BSS sequences.

Asteroseismology can provide precise constraints on the formation and evolution of BSSs. 
Although pulsating BSSs have been identified, seismic modelling accounting for scenarios beyond single-star evolution has not yet been carried out.
Theoretical work has addressed seismic signatures of interactions \citep[e.g.,][]{Rui2021,Henneco2024,Briganti2025,Miszuda2025}. 
In our previous paper \citep[hereafter DD2025]{JDD2025}, we constructed seismic models of five SX Phe stars in the core of the globular cluster $\omega$ Centauri assuming single-star evolution. This is a useful first approximation because BSSs can evolve similarly to single stars of the same mass \citep[e.g.,][]{Lombardi1996,Sills2001,Glebbeek2013}.
In this paper, we assume a collision-formation scenario and, for one star, also consider binary evolution.

In Section\,2, we describe the models and in Section\,3 the results of the seismic modelling. 
Conclusions are in Section\,4 and the conditions for rejuvenation are discussed in the Appendix.

\section{Collisional models}
To compute the collision models, we followed the approach of \citet{PortegiesZwart2019}.
The modelling scheme is illustrated in Fig.\,1. First, we evolved two main-sequence stars using MESA \citep{Paxton2013, Paxton2015,Paxton2018,Paxton2019,Jermyn2023} and subsequently merged them using
Make-Me-A-Massive-Star (MMAMS; \cite{Gaburov2008}).

MMAMS  is a 1D code that does not simulate the collision dynamics directly but instead reconstructs the merger remnant in hydrostatic equilibrium.
It uses the entropic variable of Gaburov et al. (2008) to select the denser progenitor shell at each remnant pressure, rather than simply sorting 
by entropy. Mass loss, shock heating, and mixing follow prescriptions calibrated against SPH simulations \citep{Lombardi1996,Lombardi2002,Gaburov2008}. The remnant is in hydrostatic, but not thermal, equilibrium. 

In this work, we used the Python implementation of MMAMS, PyMMAMS \citep{Heller2026}. 
The subsequent thermal relaxation and evolution of the merger remnant up to the cluster age were 
computed with MESA. Finally, non-adiabatic stellar pulsations
were calculated using the linear pulsation code of \citet{Dziembowski1977}.
\begin{figure}[ht!]
	\centering
	\includegraphics[width=\columnwidth, height=12.7cm]{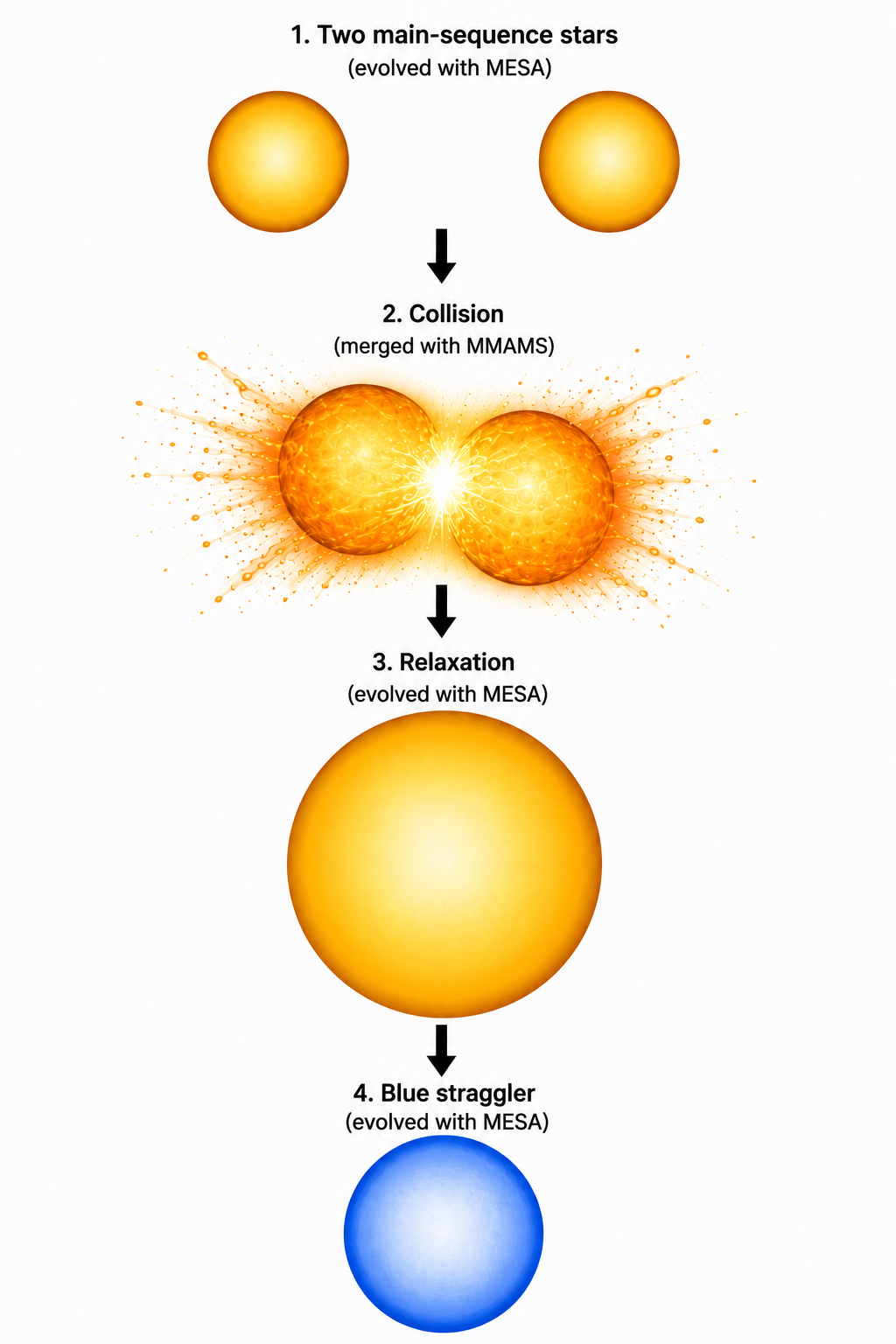}
	\caption{Schematic illustration of the collision scenario for the formation and subsequent evolution of a blue straggler star.}
	\label{fig1}
\end{figure}

In all computations, we adopted  the OPAL opacity tables  \citep{Iglesias1996} 
and the chemical mixture of \citet{Asplund2009}. 
At lower temperatures ($\log T<3.95$), we used the opacity data of \citet{Ferguson2005}. 
For the parent stars, we assumed an initial hydrogen abundance of $X_0=0.74$, which is typical of Population II.
Guided by the results of the seismic modelling of the prototype SX Phe star \citep{JDD2023}, we adopted
a mixing-length parameter of $\alpha_{\rm MLT}=1.0$ in the envelope.

\section{Seismic modelling}
We analysed the five SX Phe stars from the core of $\omega$ Cen: V194, V220, V225, V237, and NV326.
All of them pulsate in two radial modes: four in the fundamental and first-overtone modes, and one (V237)
in the fundamental and second-overtone modes. Their variability was detected by \cite{Kaluzny2004} and analysed by \citet{Olech2005}.
The globular cluster $\omega$ Cen is a very complex stellar system with a large spread in metallicity \citep[e.g.,][]{Villanova2014,Nitschai2023}. Here, we considered the metallicity range from $Z=0.001$ to $Z=0.004$.
The majority of stars in $\omega$ Cen have ages between about 10 and 13\,Gyr \citep{Clontz2024}, and we therefore considered this range.
The BSS age is counted from the end of thermal relaxation after the collision, or from the end of MT, when the orbital period becomes constant, in the binary-evolution scenario.

The stellar parameters and pulsation frequencies of the five SX Phe stars are given  in Table\,A.1. 
All five stars are located in the blue-straggler region of the CMD, as shown in Fig.\,A.1.
Seismic models were constructed by simultaneously fitting the observed frequencies 
of the two radial modes and the observed position in the Hertzsprung-Russell (HR) diagram.

\subsection{Collision-formation scenario}
For all parent stars, we assumed an initial rotational velocity of 10\,$\kms$, while for the merger remnant,
the rotational velocity was adjusted to obtain the best seismic model. All types of rotational mixing were included.
For the resulting merger, we considered models both without and with convective-core overshooting, 
adopting $f_{\rm ov}=0.02$ in the latter case. The Ledoux criterion for convective instability was adopted.
In addition, we assumed a lower limit on the chemical diffusion coefficient, adopting $\min(D_{\rm mix})=100$  and 200\,cm$^2$\,s$^{-1}$. 
Thus, in stable radiative regions where other mixing processes are absent, MESA imposes  a minimum diffusion coefficient.
The collision time $t_{\rm coll}$ was chosen such that the age of the seismic model was close to the age of the cluster.

We successfully constructed collisional seismic models for all five stars using various mixing prescriptions, $(f_{\rm ov}, D_{\rm mix})$. However, their internal structures and evolutionary stages differ significantly.
As an example, we present the results for V194 at $Z=0.003$. The left panel of Fig.\,2 shows the HR diagram
with the positions of the collisional seismic models within the error box of V194. The right panel shows the corresponding values of the central hydrogen abundance $X_{\rm cent}$ for the different combinations of $(f_{\rm ov}, D_{\rm mix})$.
The grey dots indicate the maximum values of $X_{\rm cent}$ reached by the merger remnants after the collision, while the black dots show $X_{\rm cent}$ for the corresponding collisional seismic models of V194.
\begin{figure*}[ht!]
	\centering
	\includegraphics[width=\columnwidth, height=7cm]{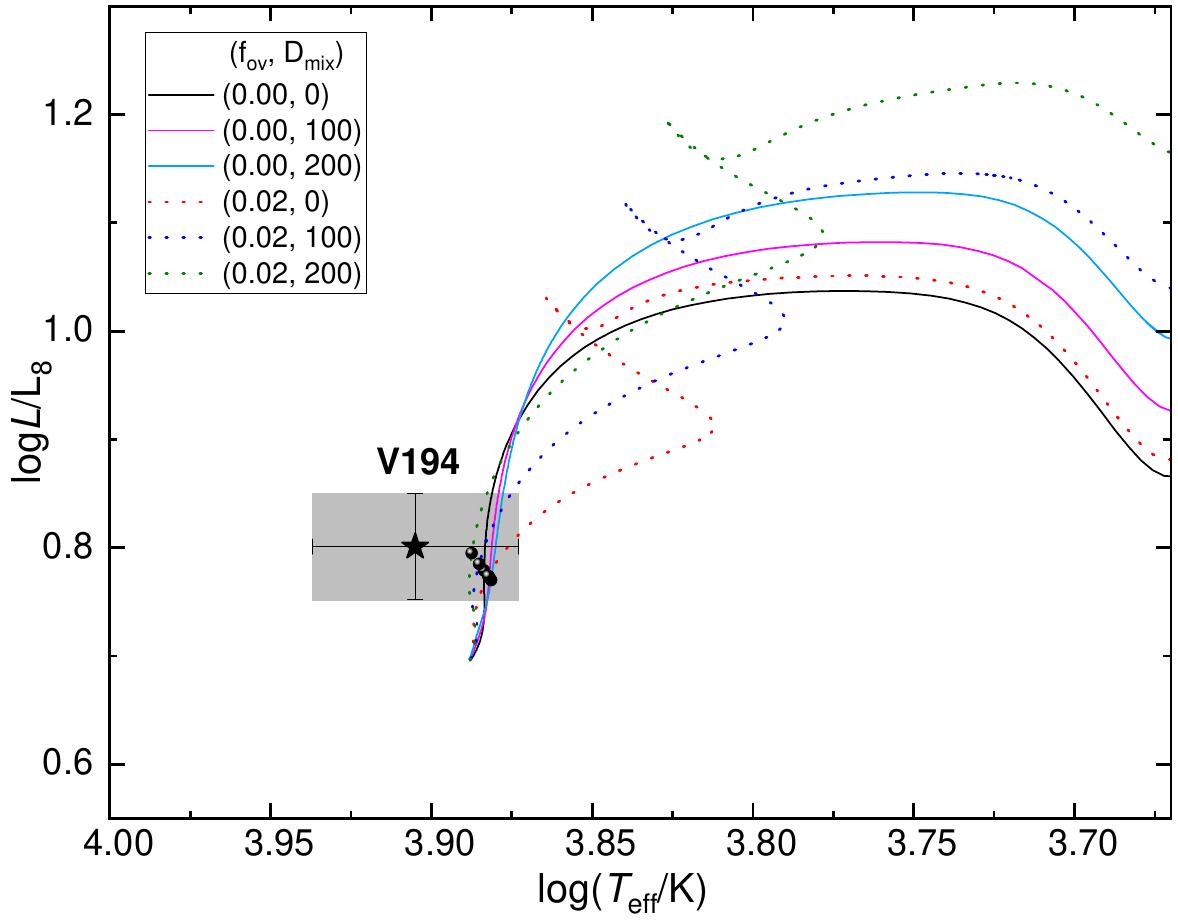}
	\includegraphics[width=\columnwidth, height=7cm]{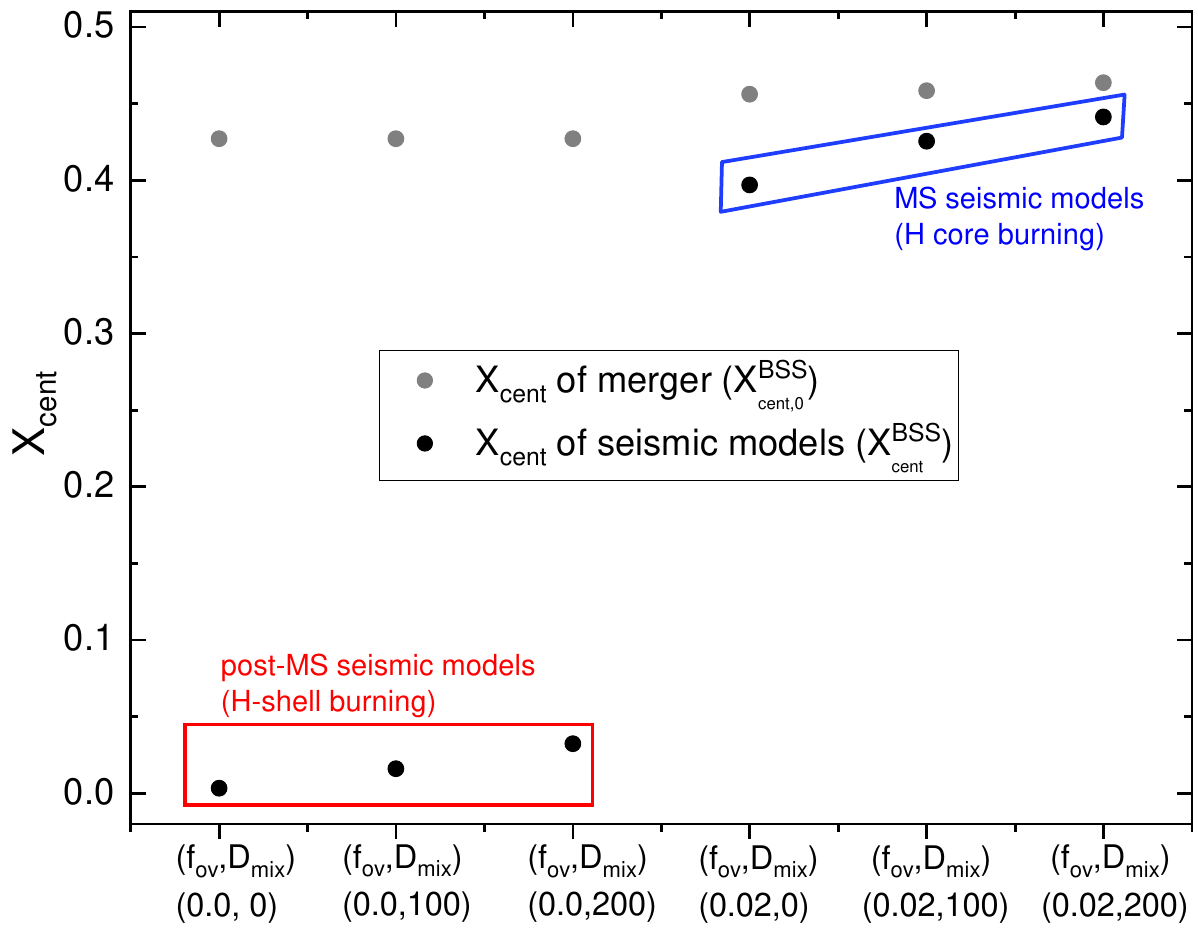}
	\caption{Left panel: Positions of the collisional seismic models of V194 in the HR diagram for a metallicity of $Z=0.003$. Right panel: Central hydrogen abundance for different mixing prescriptions. Grey dots indicate the maximum values of $X_{\rm cent}$ reached by the merger remnants after the collision, while black dots show the values of $X_{\rm cent}$ for the collisional seismic models of V194.}
	\label{fig2}
\end{figure*}
As can be seen, models including convective-core overshooting have significantly higher values of $X_{\rm cent}$ and are still in the main-sequence
phase of evolution. In contrast, models without overshooting, even when a minimum value of $D_{\rm mix}=100$ or 200 is adopted,
are already in the post-MS phase and have very little hydrogen remaining in the centre ($X_{\rm cent}\approx 1-3\% $). Thus, rejuvenation occurs only in models including convective-core overshooting.

However, convective-core overshooting is a necessary, but not always sufficient, condition for rejuvenation.
As shown in Fig.\,B.1, which presents the evolution of various parameters after  the collision,
three different cases can be distinguished: no rejuvenation (left panels), delayed rejuvenation (middle panels) and immediate rejuvenation (right panels).
The conditions that determine whether or not a merger undergoes rejuvenation are briefly discussed in Appendix\,B.
A detailed analysis will be presented by Góra et al. (in preparation).

A summary of the central hydrogen abundances of the collisional seismic models for all five stars is presented in Fig.\,C.1.
For some stars, seismic models could not be obtained for all considered metallicities. Nevertheless, a clear trend emerges:
the higher the metallicity, the more likely the merger remnant is to undergo rejuvenation, provided that convective-core overshooting is included.

The parameters of the seismic models for all stars, computed for different metallicities $Z$ and mixing prescriptions $(f_{\rm ov}, D_{\rm mix})$ are given in Table\,C.1. For the remaining four stars, we considered the following mixing prescriptions: $(f_{\rm ov},D_{\rm mix})=(0.00,~0)$, $(0.02,~0)$, and $(0.02,~200)$.

Fig.\,3 shows the dependence of the seismic masses on metallicity for all five stars. The masses are the same for all pairs 
of $(f_{\rm ov}, D_{\rm mix})$. A clear trend is seen: the higher the metallicity, 
the higher the seismic mass. For comparison, the open circles indicate the seismic masses obtained in our previous study (DD2025) assuming
single-star evolution. The masses of the collisional seismic models are systematically higher than those obtained from single-star evolution.
However, this comparison should be treated with caution because convective-core overshooting was not included in the single-star evolutionary models.

Fig.\,4 shows the ages of the seismic models for all five stars. The four upper panels, from top to bottom, correspond to $Z=0.001$, 0.002, 0.003 and 0.004,
respectively. The bottom panel shows the ages of the seismic models obtained by DD2025 assuming single-star evolution.
We find the rather intuitive result that additional mixing slightly increases the age of the seismic model. Comparing, again with caution, the collisional seismic models with those obtained assuming single-star evolution, we find that the latter are systematically older.
 
\begin{figure}[ht!]
	\centering
	\includegraphics[width=\columnwidth,height=5.1cm]{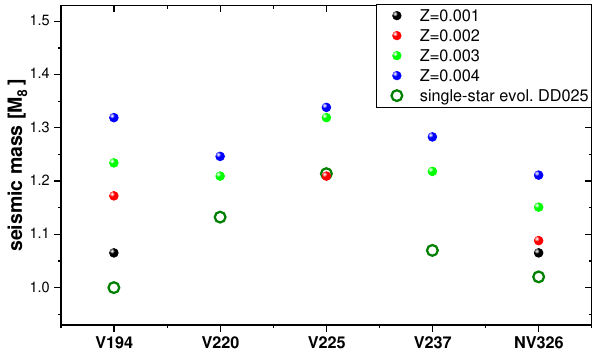}
	\caption{Masses of the collisional seismic models of the five SX Phe stars for four values of metallicity $Z$. Open circles indicate the seismic masses obtained assuming single-star evolution.}
	\label{fig3}
\end{figure}

\begin{figure}[ht!]
	\centering
	\includegraphics[width=\columnwidth]{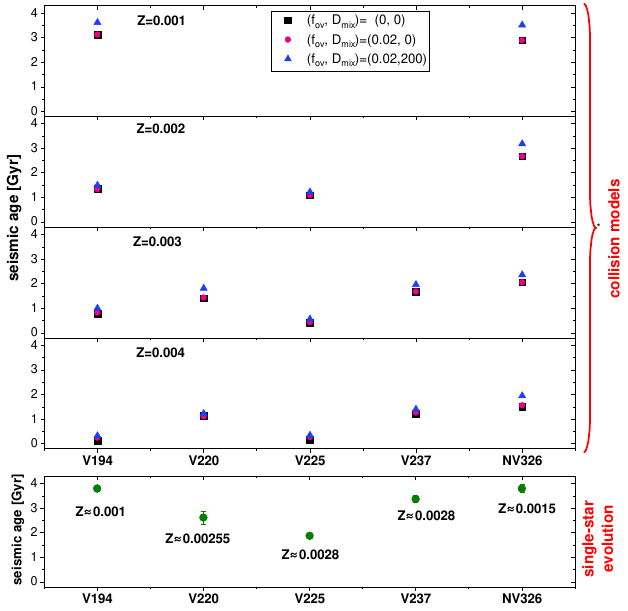}
	\caption{Four upper panels: ages of collisional seismic models for the five SX Phe stars for four values of $Z$. Bottom panel: ages of the seismic models obtained assuming single-star evolution.}
	\label{fig4}
\end{figure}

\subsection{The binary-evolution scenario for the star V237}
Among the five stars studied, V237 is located farthest from the centre of $\omega$ Cen.
To test whether the available seismic and $(T_{\rm eff},~L)$ constraints can discriminate between formation
channels, we also considered binary evolution through MT for this star.
We used the MESA-Binary module to construct a grid of seismic models that simultaneously reproduce the observed frequencies of the radial fundamental and second-overtone modes. 

All models were computed using the microphysics described in Sect.\,2. We adopted convective-core overshooting with $f_{\rm ov}=0.02$, $\min(D_{\rm mix})=10$\,cm$^2$\,s$^{-1}$, and $V_{\rm rot}=0\,\kms$. Over a range of metallicities, $Z\in(0.001,0.004)$, initial orbital periods of  $0.5 - 5$\,d, and assuming the circular orbit ($e=0$), we found three seismic models, all with an initial orbital period of 2\,d and all with conservative MT. Their parameters are listed in Table\,1. Fig.\,5 shows that none lies within the observed error box in the HR diagram: model 1 lies near the boundary of the 3$\sigma$ error, while models 2 and 3 are outside it. 
Thus, although both formation channels reproduce the two radial-mode frequencies, the stellar parameters favour a collisional origin for V237.
\begin{table}[ht!]
	\caption{Parameters of the binary seismic models of V237. }
	\centering
	\begin{tabular}{cccccccc}
		\hline
		No. & $(M_1,~M_2)$  & $P_\mathrm{orb}$ & $M_{\rm BSS}$ &  $Z$  & $X_{\rm cent}^{\rm BSS}$ &  age \\
		&  [M$_{\sun}$] & [d] & [M$_{\sun}$] &  &  &  [Gyr] &  \\
		\hline
		1 & (1.0, 0.6) & ~~8.04  &1.124 & 0.003 & 0.18$^1$ &  3.23    \\
		2 & (0.9, 0.7) &   10.89 &1.355 & 0.001 & 0.01$^2$ &  1.50   \\
		3 & (1.1, 0.6) &   ~~9.69 &1.066 & 0.003 & 0.17$^1$ &   4.13   \\
		\hline
	\end{tabular}
	\tablefoot{The columns give the masses of the donor and accretor, the current orbital period, the BSS mass, metallicity, central hydrogen abundance, and age of the seismic model. 1 -- MS stage, 2 -- post-MS stage.}
\end{table}
\begin{figure}[ht!]
	\centering
	\includegraphics[width=\columnwidth,height=5.0cm]{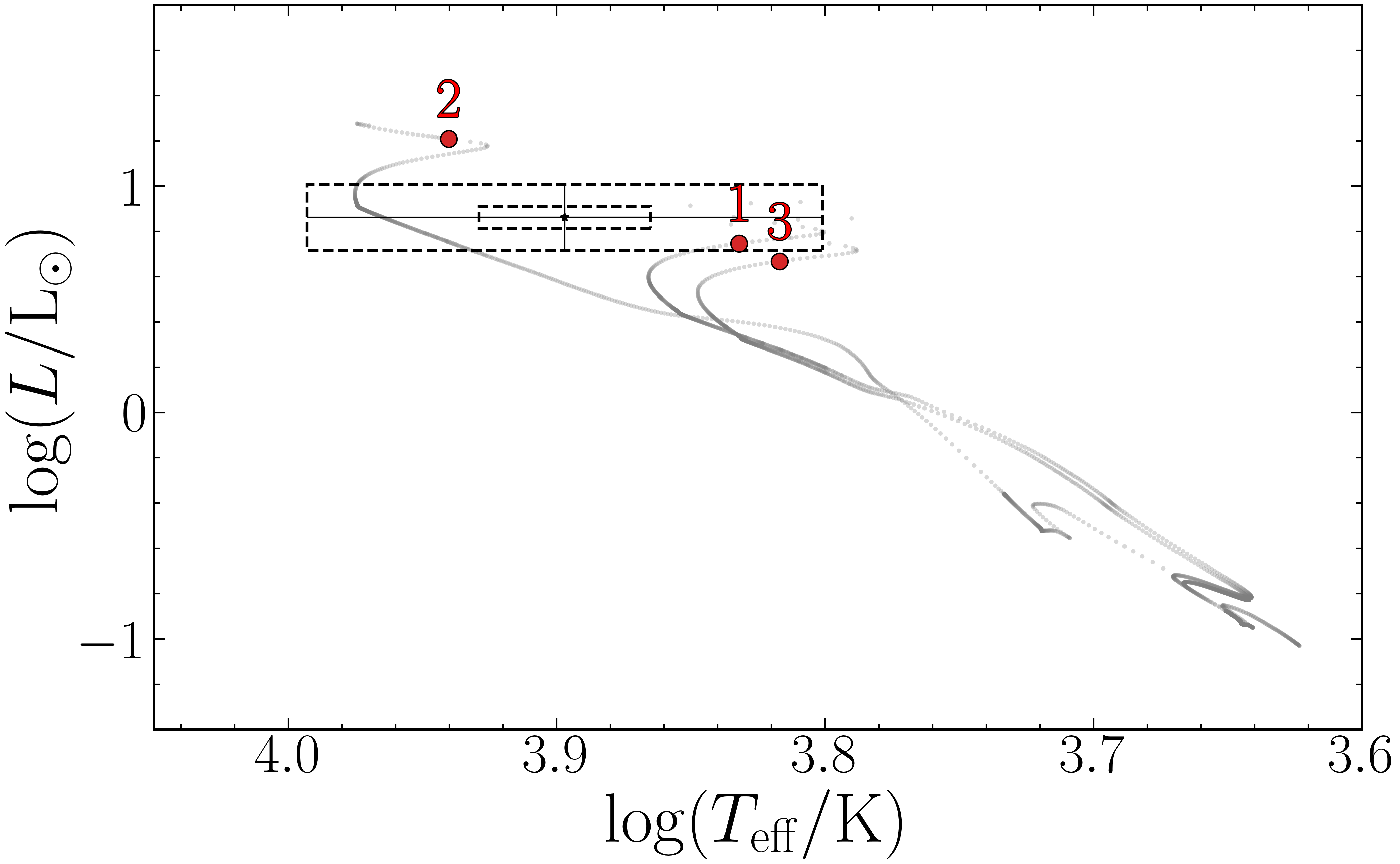}
	\caption{Positions of the seismic models of V237 in the HR diagram obtained under the binary-evolution scenario with MT. The two error boxes indicate the $1\sigma$ and $3\sigma$ uncertainties.}
\label{figC2}
\end{figure}

\section{Conclusions}
We presented the first seismic modelling of pulsating blue stragglers under the assumption of
a collision-formation scenario. To this end, we selected five SX Phe stars from the globular cluster
$\omega$ Cen, all lying in the BSS region of the CMD. All five stars are located in 
the cluster core, lending support to their possible formation through collisions.
Collisional seismic models were obtained for all five SX Phe stars, yielding systematically higher masses and younger ages
than the single-star evolution models. For V237, located farthest from the cluster centre, three binary-MT models 
also reproduce the two radial-mode frequencies, but not the observed position in the HR diagram. 
Thus, within our model grids, a collisional origin is favoured.

The internal structure and evolutionary stage of the collisional seismic models  strongly depend on the adopted mixing prescription. In particular, the inclusion of convective-core overshooting is a necessary, but not always sufficient, condition for post-collision rejuvenation. The critical factor is whether and when a convective core develops after the collision, allowing overshooting to enhance the mixing and promote further growth of the convective core. 

We identify three qualitatively different responses of the merger remnants to overshooting, resulting in no, delayed, or immediate rejuvenation.
Imposing the minimum diffusion coefficient, $\min(D_{\rm mix})$ does not lead to rejuvenation, 
but by continuously supplying hydrogen to the central regions slows down central hydrogen depletion 
and extends the main-sequence phase of the merger remnant.
In some models adding  $\min(D_{\rm mix})$ can delay the development of a substantial convective core and hence the onset of rejuvenation. 
The answer to the question ``to rejuvenate or not to rejuvenate'' is a complex function of the parent-star masses, metallicity, and collision time.

\begin{acknowledgements}
  The work was financially supported by the Polish National Science Centre grant 2023/50/A/ST9/00144,
  ERC (EU Horizon 2020, grant 945806), and DFG (Germany's Excellence Strategy, EXC 2181/1-390900948; Heidelberg STRUCTURES). F.R.N.S. and M.H. acknowledge support from the Klaus Tschira Foundation.
  Computations used WCSS resources, grant no. 265. 
\end{acknowledgements}

  \bibliographystyle{aa} 
  \bibliography{JDD_biblio_aa} 

\begin{appendix}
\nolinenumbers

\section{The five SX Phe stars in $\omega$ Centauri}
Table\,A.1 summarizes the values of the effective temperature and luminosity adopted in this paper as determined by DD2025.
For three stars, projected rotational velocities are also available from the literature. In the right part of Table\,A.1, we list the pairs
of frequencies, including the dominant one, whose frequency ratios are consistent with their identification
as radial modes.
In Fig.\,A.1, we show the positions of the five SX Phe stars in two CMDs; one based on the Johnson $BV$ passbands adopted from \citet{Olech2005}
and the other on the Gaia $BR$ photometry. All five stars are located in the region of blue stragglers. 
\begin{strip}
	\centering	
	
		\captionof{table}{Observed stellar parameters and pulsation frequencies of the five SX Phe stars in $\omega$ Cen.}	
	\centering
	\label{tab:TL_5stars}
	\begin{tabular}{cccc|cccc}
		\hline
		Star & $\log(T_{\rm eff}/\rm K)$ & $\log(L/\rm L_{\odot})$ & $V_{\rm rot}\sin i$ &   $\nu_1$ &   $\nu_n$ & $\nu_1/\nu_n$&    pulsation \\
		&                 &         &      $[\kms]$   & $[\mathrm d^{-1}]$ & $[\mathrm d^{-1}]$ & & modes\\ 
		\hline
		V194 &  3.905(32) & 0.801(49) & -- 	&  21.1964(1) & 27.1633(5) & 0.78033(2) & \bf{F+1O}\\
		V220 &  3.894(32) & 0.811(48) & 22$^1$ & 18.9083(2) & 24.3161(5) &  0.77760(2) & \bf{F+1O} \\
		V225 &  3.879(32) & 0.871(51) & 20$^1$ &  20.5600(2) &  26.4145(5) & 0.77836(2) & \bf{F+1O}\\
		V237 &  3.897(32) & 0.861(48) & -- & 15.2433(2) & 24.1162(5) &  0.63208(2) & \bf{F+2O}\\
		NV326 &  3.876(32) & 0.793(51) & $55.9\pm24.7^2$ & 17.5728(6) & 22.5375(6) & 0.77971(3) & \bf{F+1O}\\
		\hline
	\end{tabular}\\
    \tablefoot{Left: Effective temperature, luminosity, and projected rotational velocities from  (1) \citet{Mucciarelli2014} and (2)~~ \citet{Simunovic2014}. Right: Dominant frequency and the frequency yielding a frequency ratio indicative of radial modes, with the corresponding mode identification given in the last column.}

\vspace{0.25cm}

	\captionof{figure}{CMD of $\omega$ Cen in the Johnson $BV$ passbands, adapted from \citet{Olech2005} (left panel), and CMD based on Gaia photometry (right panel). The five analysed SX Phe stars are marked in colour. Open circles in the left panel indicate other SX Phe stars in $\omega$ Cen, while points denote constant stars. All five analysed stars lie in the blue-straggler region.}
	\includegraphics[width=9.0cm, height=10.0cm]{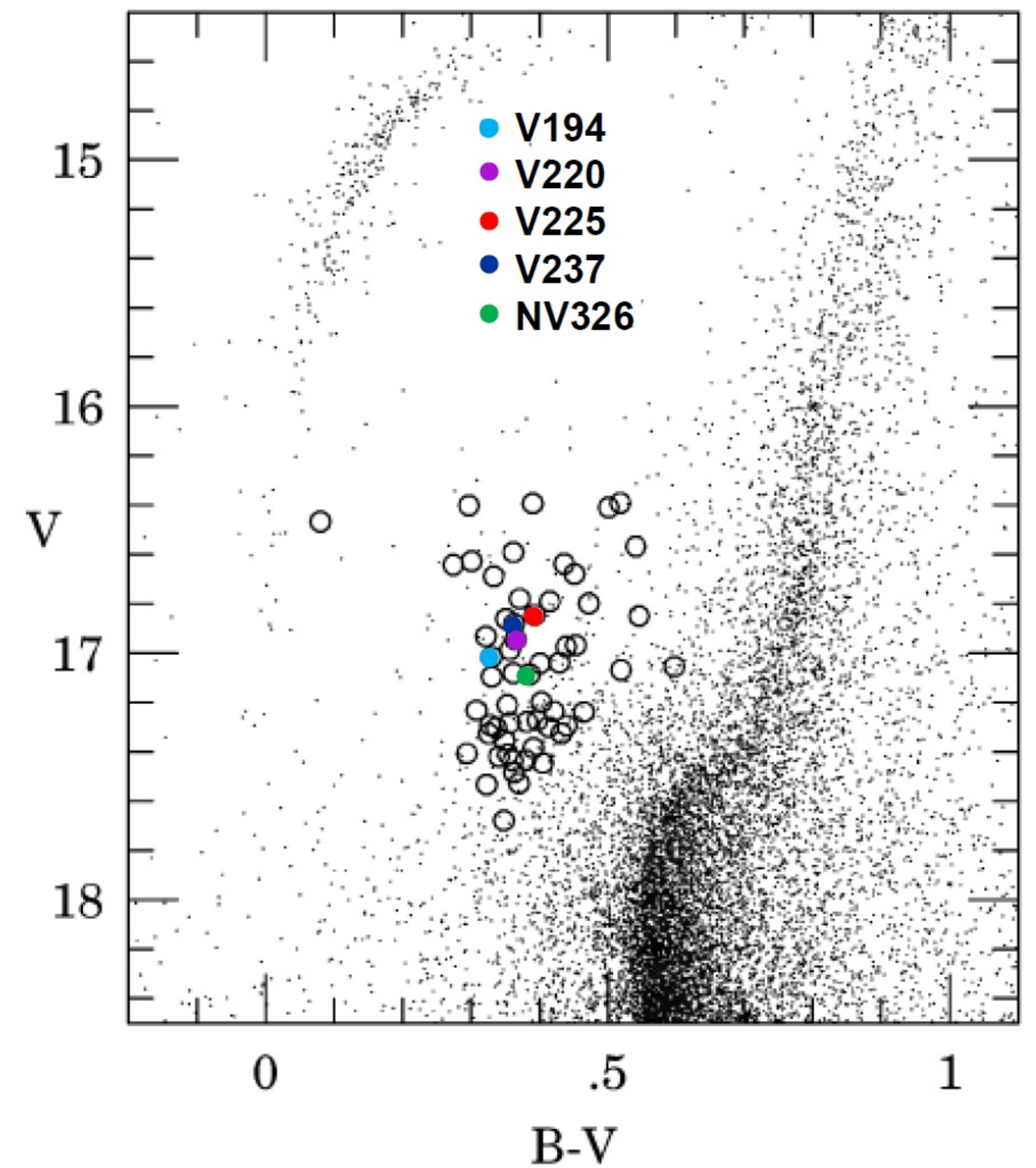}
	\includegraphics[width=9.0cm, height=10cm]{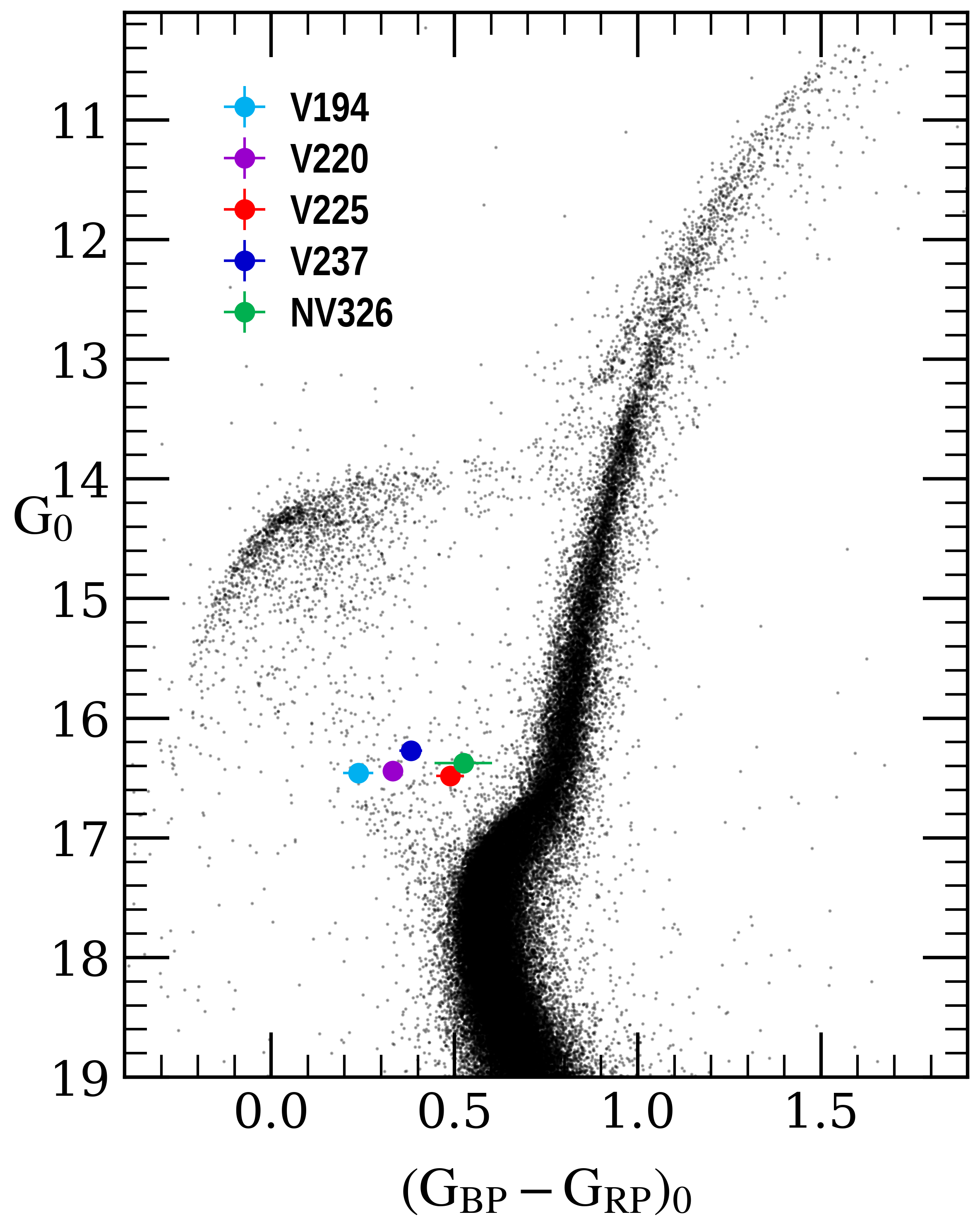}
	\label{figA1}

\end{strip}

\section{To rejuvenate or not to rejuvenate}
During the seismic modelling of the five SX Phe stars in $\omega$ Cen under the collision-formation scenario,  we identified three distinct evolutionary behaviours of the merger remnants: no rejuvenation, delayed rejuvenation, and immediate rejuvenation.
Representative examples are shown in Fig.\,B.1. From top to bottom, the panels show the evolution of the central hydrogen abundance; the convective-core mass and central mean molecular weight; the central nuclear energy-generation rates from the pp and CNO cycles; and the central values of  temperature and density. Three combinations of $(f_{\rm ov},D_{\rm mix})$ are shown: (0, 0), (0.02, 0) and (0.02, 200). The models have the same metallicity, $Z=0.003$, but differ in their parent-star masses and collision times. 
The left panels show the V237 model, representing the no-rejuvenation case; the middle panels show the V194 model, illustrating delayed rejuvenation; and the right panels show the V225 model, illustrating immediate rejuvenation. These three cases represent qualitatively different responses of the merger remnants to convective-core overshooting.\\
 
\noindent {\bf No rejuvenation}\\
In the left panels (the V237 model), $X_{\rm cent}$ decreases monotonically for each combination of $(f_{\rm ov},D_{\rm mix})$, while the second panel from the top shows that $M_{\rm cc}\approx0$. The post-merger core never fulfils the Ledoux criterion for convective instability. 
The central energy-generation rate from the pp chain, $\varepsilon_{\rm pp}$, decreases monotonically, whereas the central value of $\varepsilon_{\rm CNO}$ initially increases due to the increase in the central temperature and then decreases at a similar rate.
Thus, the central hydrogen is depleted rather rapidly, the central mean molecular weight increases and no rejuvenation takes place. Adopting $D_{\rm mix}=200$\,cm$^2$\,s$^{-1}$ considerably prolongs the central hydrogen-depletion time, but does not lead to the formation of a convective core.\\

\noindent {\bf Delayed rejuvenation}\\
The evolution of parameters in the middle panels for the model of V194 is fundamentally different.
Immediately after the collision, $X_{\rm cent}$ decreases in all models. Thus, the V194 model initially behaves like the V237 model. 
For the overshooting model, however, the structure changes after a few $10^8$ yr,
because a small convective core starts to develop. This occurs at the moment when the decline of $X_{\rm cent}$ stops and reverses. 
This happens because convective-core overshooting becomes effective. It mixes across 
the core boundary into H-richer material, so  $X_{\rm cent}$  increases and $\mu_c$ decreases.

Before rejuvenation, the central  $\varepsilon_{\rm pp}$ and $\varepsilon_{\rm CNO}$ contributions evolve similarly to those in the V237 model. 
Around the onset of convective-core growth, the central nuclear energy rate changes, 
and after rejuvenation the pp contribution remains high and nearly constant for a long time.
This suggests that thermal and nuclear evolution before rejuvenation eventually makes the central region convectively unstable. Once the convective core appears, convective-core overshooting enhances chemical mixing, leading to an increase in $X_{\rm cent}$.\\

\noindent {\bf Immediate rejuvenation}\\
The model shown in the right panels exhibits yet another evolutionary behaviour. 
Immediately after the collision, $X_{\rm cent}\approx 0.3$ and in both models with overshooting 
it rises almost instantaneously to $ X_{\rm cent} \approx 0.48.$
At the same time, the second panel shows that the mass of the convective core  
increases to more than 10\% of the total mass, while $\mu_c$ simultaneously decreases.
This indicates that the post-merger structure of the V225 model is very close to satisfying the Ledoux criterion for convective instability from the very beginning and satisfies it shortly thereafter.

The panel showing the nuclear generation rate is also revealing. The immediate rejuvenation is not preceded by a phase of increasing $\varepsilon$
from CNO burning. The central contribution of $\varepsilon$ from pp burning becomes large and remains large after fresh H is supplied.\\

These three cases suggest that convective-core overshooting is necessary for rejuvenation, but does not entirely determine whether
a convective core will develop. Thus, overshooting acts rather as an amplifier of convective-core development. The V237 model never activates
this feedback, V194 activates it only after some thermal and nuclear evolution, whereas V225 activates it almost immediately after the collision. 
As already mentioned, this issue, including a detailed analysis of the internal structure of post-merger products, will be investigated in a separate paper (Góra et al., in preparation).
\begin{figure*}[ht!]
	\centering
	\includegraphics[width=0.33\textwidth,height=13.9cm]{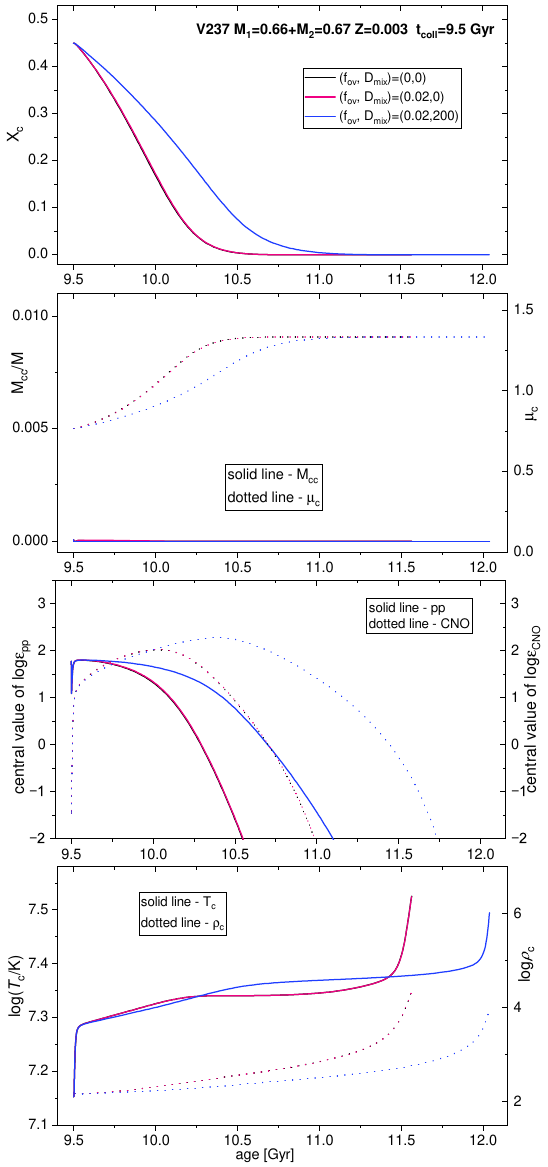}
	\includegraphics[width=0.33\textwidth,height=13.9cm]{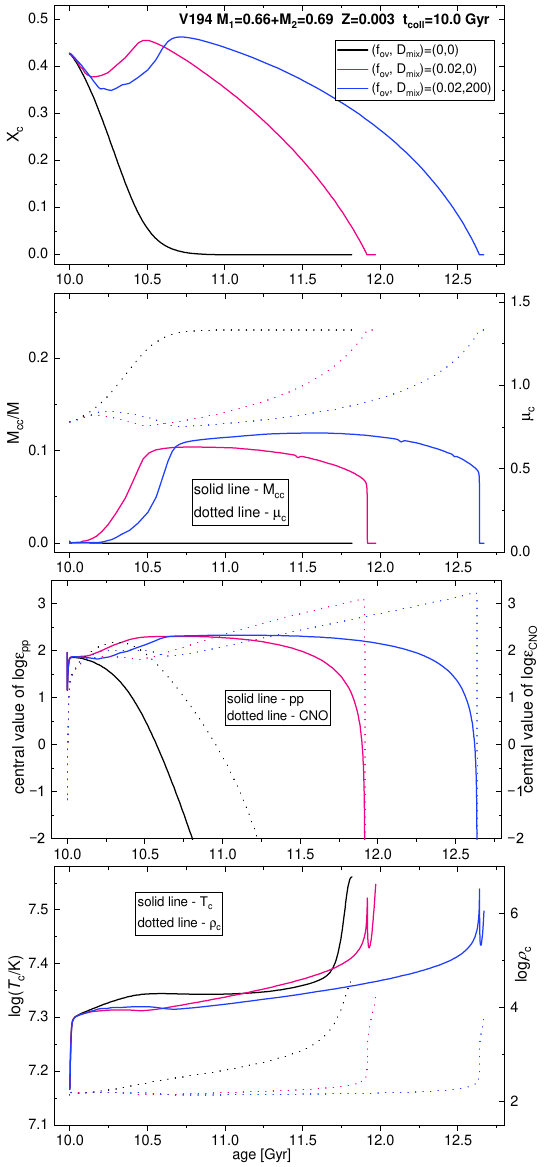}
	\includegraphics[width=0.33\textwidth,height=13.9cm]{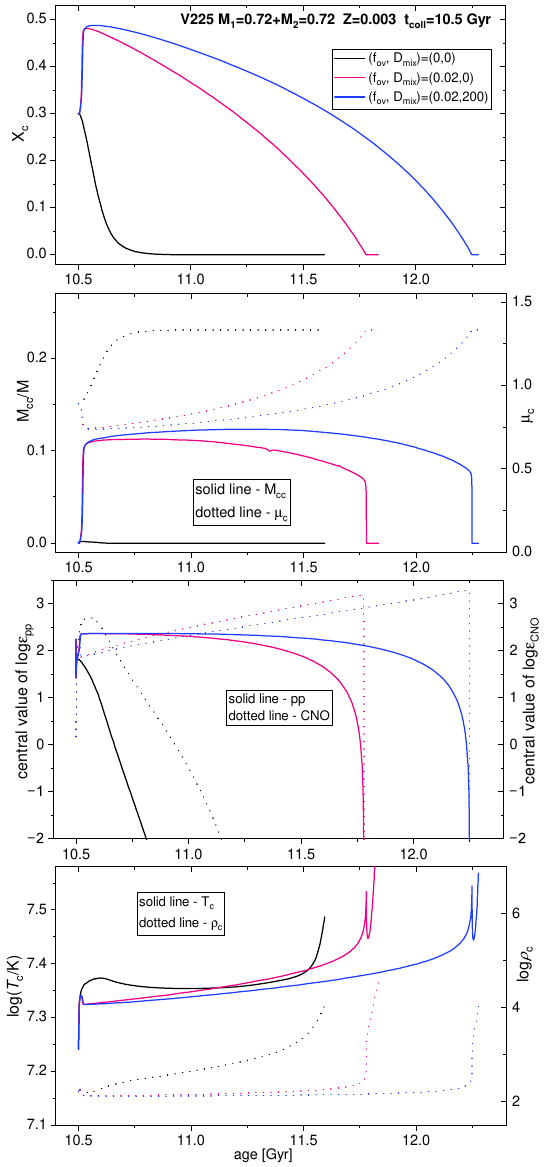} 
	\caption{Evolution of three representative post-collision models illustrating the different responses to convective-core overshooting. From left to right: V237, V194, and V225, representing no, delayed, and immediate rejuvenation, respectively. From top to bottom, the panels show the central hydrogen abundance, $X_{\rm cent}$; the convective-core mass, $M_{\rm cc}/M$ (solid lines), and the central mean molecular weight, $\mu_{\rm c}$ (dotted lines); the central energy-generation rates from the pp chains (solid lines) and CNO cycle (dotted lines); and the central temperature, $T_{\rm c}$ (solid lines), and density, $\rho_{\rm c}$ (dotted lines). Black, magenta, and blue lines correspond to 
		$(f_{\rm ov},~D_{\rm mix})=(0,0),~  (0.02,0)$, and $(0.02,200 \,{\rm cm^2\,s^{-1}})$, respectively. The three models have the same metallicity, \(Z=0.003\), but differ in the parent-star masses and collision times, as indicated in the top panels.}
	\label{figB1}
\end{figure*}

\section{Parameters of collisional seismic models}
Table\,C.1 summarizes the parameters of the seismic models obtained under the collision-formation scenario for the five SX Phe stars
in the globular cluster $\omega$ Cen. All models reproduce the observed frequencies of two radial modes. 
The second column lists the two mixing parameters, $f_{\rm ov}$ and $D_{\rm mix}$, which specify the extent of convective-core overshooting and the minimum value of the total diffusive coefficient, respectively. 
The third column gives the metallicity $Z$,  assumed to be the same for both parent stars. The next three columns give the masses of the parent stars $(M_1,~M_2)$ and the mass of the collisional seismic model $M_{\rm BSS}$. The seventh column lists the collision time. The subsequent columns give the effective temperature, luminosity, rotational velocity, and age of the seismic model
measured from the time of collision. 
The last two columns contain the maximum central hydrogen abundance reached by the merger remnant after the collision 
and the central hydrogen abundance of the collisional seismic model, respectively. In all cases, an initial hydrogen abundance of $X_0=0.74$, typical of Population II stars, was assumed for the parent stars.
\begin{table*}[ht!]
	\caption{Parameters of the collisional seismic models of the five SX Phe stars in $\omega$ Cen.}
	\centering
	\begin{tabular}{ccccccccccccccc}
		\hline
		Star & $(f_{\rm ov},~D_{\rm mix})$& $Z$ & $M_1$& $M_2$& $M_{\rm BS}$& $t_{\rm coll}$& $\log (T_{\rm eff}/{\rm K})$& $\log L/{\rm L}_{\odot}$& $V_{\rm rot}^{\rm BSS}$& age$_{\rm BSS}$& $X_{{\rm cent},0}^{\rm BSS}$ & $X_{\rm cent}^{\rm BSS}$ \\
		& & & [M$_{\odot}$] & [M$_{\odot}$] & [M$_{\odot}$] & [Gyr] & & & $[\kms]$ & [Gyr] & & \\
		\hline
		\hline
		V194 & (0.00, 0.0) & 0.001 & 0.60 & 0.565 & 1.065 & 9.0 & 3.8872 & 0.756 & 57 & 3.12 & 0.5594 & 0.0001 \\
	         & (0.02, 0.0) & 0.001 & 0.60 & 0.565 & 1.065 & 9.0 & 3.8869 & 0.755 & 41 & 3.12 & 0.5638 & 0.0001  \\
		& (0.02, 200) & 0.001 & 0.60 & 0.565 & 1.065 & 9.0 & 3.8904 & 0.769 & 46 & 3.61 & 0.5647 & 0.0110 \\
		\cline{2-13}
		& (0.00, 0.0) & 0.002 & 0.64 & 0.64 & 1.172 & 10.0 & 3.8872 & 0.781 & 65 & 1.33 & 0.4393 & 0.0005 \\
		& (0.02, 0.0) & 0.002 & 0.64 & 0.64 & 1.172 & 10.0 & 3.8872 & 0.781 & 65 & 1.33 & 0.4393 & 0.0006 \\
		& (0.02, 200) & 0.002 & 0.64 & 0.64 & 1.172 & 10.0 & 3.8853 & 0.773 & 62 & 1.49 & 0.4393 & 0.0204 \\
		\cline{2-13}
		& (0.00, 0.0) & 0.003 & 0.66 & 0.69 & 1.234 & 10.0 & 3.8836 & 0.779 & 71 & 0.78 & 0.4271 & 0.0033 \\
		& (0.00, 100) & 0.003 & 0.66 & 0.69 & 1.234 & 10.0 & 3.8820 & 0.773 & 68 & 0.81 & 0.4271 & 0.0161 \\
		& (0.00, 200) & 0.003 & 0.66 & 0.69 & 1.234 & 10.0 & 3.8814 & 0.770 & 70 & 0.87 & 0.4271 & 0.0324 \\
		& (0.02, 0.0) & 0.003 & 0.66 & 0.69 & 1.234 & 10.0 & 3.8823 & 0.774 & 70 & 0.83 & 0.4562 & 0.3970 \\
		& (0.02, 100) & 0.003 & 0.66 & 0.69 & 1.234 & 10.0 & 3.8851 & 0.785 & 73 & 0.92 & 0.4584 & 0.4255 \\
		& (0.02, 200) & 0.003 & 0.66 & 0.69 & 1.234 & 10.0 & 3.8874 & 0.795 & 76 & 1.00 & 0.4636 & 0.4414 \\
		\cline{2-13}
		& (0.00, 0.0) & 0.004 & 0.72 & 0.72 & 1.319 & 11.0 & 3.8830 & 0.794 & 70 & 0.11 & 0.3170 & 0.1136 \\
		& (0.02, 0.0) & 0.004 & 0.72 & 0.72 & 1.319 & 11.0 & 3.8902 & 0.823 & 82 & 0.25 & 0.4965 & 0.4593 \\
		& (0.02, 200) & 0.004 & 0.72 & 0.72 & 1.319 & 11.0 & 3.8924 & 0.832 & 84 & 0.32 & 0.5040 & 0.4845 \\
		\hline
		\hline
		V220 & (0.00, 0.0) &0.003 & 0.66 & 0.66 & 1.209 & 9.5 &3.8758 & 0.808 & 36 & 1.40 & 0.4481 & 0.0000 \\
		& (0.02, 0.0) &0.003 & 0.66 & 0.66 & 1.209 & 9.5 &3.8677 & 0.775 & 14 & 1.44 & 0.5711 & 0.3200 \\
		& (0.02, 200) &0.003 & 0.66 & 0.66 & 1.209 & 9.5 &3.8799 & 0.825 & 33 & 1.81 & 0.5788 & 0.3966 \\
		\cline{2-13}
		& (0.00, 0.0) &0.004 & 0.68 & 0.68 & 1.246 & 10.0 &3.8661 & 0.776 & 48 & 1.13 & 0.4432 & 0.0003 \\
		& (0.02, 0.0) &0.004 & 0.68 & 0.68 & 1.246 & 9.5 &3.8620 & 0.760 & 41 & 1.11 & 0.5121 & 0.3721 \\
		& (0.02, 200) &0.004 & 0.68 & 0.68 & 1.246 & 10.0 &3.8626 & 0.762 & 41 & 1.22 & 0.4432 & 0.0134 \\
		\hline
		\hline
		V225 & (0.00, 0.0) &0.002 & 0.66 & 0.66 & 1.209 & 10.5 &3.8989 & 0.854 & 35 & 1.08 &0.3871 & 0.0001\\
		& (0.02, 0.0) &0.002 & 0.66 & 0.66 & 1.209 & 10.5 &3.8990 & 0.854 & 34 & 1.09 &0.3871 & 0.0001\\
		& (0.02, 200) &0.002 & 0.66 & 0.66 & 1.209 & 10.5 &3.8964 & 0.844 & 26 & 1.21 &0.3871 & 0.0050\\
		\cline{2-13}
		& (0.00, 0.0) &0.003 & 0.72 & 0.72 & 1.319 & 10.5 &3.9067 & 0.908 & 43 & 0.40 &0.3007 & 0.0005\\
		& (0.02, 0.0) &0.003 & 0.72 & 0.72 & 1.319 & 10.5 &3.9050 & 0.901 & 38 & 0.44 &0.4816 & 0.3846\\
		& (0.02, 200) &0.003 & 0.72 & 0.72 & 1.319 & 10.5 &3.9098 & 0.921 & 48 & 0.56 &0.4882 & 0.4266\\
		\cline{2-13}
		& (0.00, 0.0) &0.004 & 0.73 & 0.73 & 1.338 & 11.0 &3.8887 & 0.838 & 38 & 0.16 &0.3112 & 0.0294\\
		& (0.02, 0.0) &0.004 & 0.73 & 0.73 & 1.338 & 11.0 &3.8946 & 0.862 & 55 & 0.27 &0.4822 & 0.4354\\
		& (0.02, 200) &0.004 & 0.73 & 0.73 & 1.338 & 11.0 &3.8972 & 0.873 & 59 & 0.34 &0.4900 & 0.4639\\
		\hline
		\hline
		V237 & (0.00, 0.0) &0.003 & 0.66 & 0.67 & 1.218 & 9.5 &3.8677 & 0.902 & 59 & 1.68 & 0.4504& 0.0000\\
        	& (0.02, 0.0) &0.003 & 0.66 & 0.67 & 1.218 & 9.5 &3.8677 & 0.902 & 59 & 1.68 & 0.4504& 0.0000\\
		& (0.02, 200) &0.003 & 0.66 & 0.67 & 1.218 & 9.5 &3.8682 & 0.904 & 59 & 1.96 & 0.4504& 0.0001\\
		\cline{2-13}
		& (0.00, 0.0) &0.004 & 0.70 & 0.70 & 1.283 & 10.0 &3.8662 & 0.909 & 75 & 1.22 & 0.3933& 0.0000\\
		& (0.02, 0.0) &0.004 & 0.70 & 0.70 & 1.283 & 10.0 &3.8525 & 0.854 & 60 & 1.26 & 0.4668& 0.2647\\
		& (0.02, 200) &0.004 & 0.70 & 0.70 & 1.283 & 10.0 &3.8643 & 0.901 & 73 & 1.39 & 0.3933& 0.0001\\
		\hline
		\hline
		NV326 & (0.00, 0.0) &0.001 & 0.60 & 0.565& 1.065 & 9.0 &3.8805 & 0.838 & 19 & 2.89 & 0.5610& 0.0000\\
		& (0.02, 0.0) &0.001 & 0.60 & 0.565& 1.065 & 9.0 &3.8804 & 0.838 & 17 & 2.89 & 0.5664& 0.0000\\
		& (0.02, 200) &0.001 & 0.60 & 0.565& 1.065 & 9.0 &3.8887 & 0.871 & 33 & 3.51 & 0.5674& 0.0002\\
		\cline{2-13}
		& (0.00, 0.0) &0.002 & 0.61 & 0.58 & 1.088 & 9.5 &3.8565 & 0.745 & 56 & 2.66 & 0.5537& 0.0000\\
		& (0.02, 0.0) &0.002 & 0.61 & 0.58 & 1.088 & 9.5 &3.8564 & 0.745 & 57 & 2.66 & 0.5593& 0.0000\\
		& (0.02, 200) &0.002 & 0.61 & 0.58 & 1.088 & 9.5 &3.8590 & 0.755 & 57 & 3.17 & 0.5593& 0.0029\\
		
		\cline{2-13}
		& (0.00, 0.0) &0.003 & 0.61 & 0.65 & 1.151 & 9.5 &3.8551 & 0.754 & 69 & 2.06 & 0.5236& 0.0000\\
		& (0.02, 0.0) &0.003 & 0.61 & 0.65 & 1.151 & 9.5 &3.8551 & 0.753 & 69 & 2.06 & 0.5258& 0.0000\\
		& (0.02, 200) &0.003 & 0.61 & 0.65 & 1.151 & 9.5 &3.8540 & 0.749 & 68 & 2.36 & 0.5258& 0.0047\\
		\cline{2-13}
		& (0.00, 0.0) &0.004 & 0.61 & 0.72 & 1.211 & 9.5 &3.8534 & 0.759 & 80 & 1.50 & 0.5458& 0.0001\\
		& (0.02, 0.0) &0.004 & 0.61 & 0.72 & 1.211 & 9.5 &3.8474 & 0.735 & 76 & 1.56 & 0.5967& 0.3481\\
		& (0.02, 200) &0.004 & 0.61 & 0.72 & 1.211 & 9.5 &3.8591 & 0.783 & 83 & 1.94 & 0.6022& 0.4156\\
		\hline
	\end{tabular}
	\tablefoot{The initial hydrogen abundance of the parent stars was fixed at $X_0=0.74$. The columns are described in the text.}
	\end{table*}

As mentioned above, the central hydrogen abundance of the collisional seismic models strongly depends on the adopted mixing prescription, and some merger remnants do not undergo rejuvenation. 
Fig.\,C.1 shows the dependence of the central hydrogen abundance, $X_{\rm cent}$, on metallicity and mixing for the five SX Phe  stars. The coverage in metallicity is incomplete for some stars, as collisional seismic models could not be obtained at every value of $Z$. 
Models without overshooting generally have very low values of $X_{\rm cent}$, whereas models
with $f_{\rm ov}=0.02$ can have substantially higher $X_{\rm cent}$.

At the two lowest metallicities, $Z=0.001$ and 0.002, the collisional seismic models have very little hydrogen left in their centres, irrespective of the adopted mixing prescription. At $Z=0.003$, models with convective-core overshooting reach substantially higher values of $X_{\rm cent}$ for three stars, and adopting  min$(D_{\rm mix})$=200\,cm$^2$\,s$^{-1}$ leads to a further slight increase in $X_{\rm cent}$.  At the highest metallicity considered, $Z=0.004$, all five stars undergo rejuvenation.
This behaviour suggests that metallicity, together with the efficiency of internal mixing, plays a decisive  role in determining the post-collision evolutionary state of the merger remnant.
Our results indicate that rejuvenation becomes more likely with increasing metallicity, 
provided that convective-core overshooting is included.
\begin{figure}[ht!]
	\centering
	\includegraphics[width=\columnwidth]{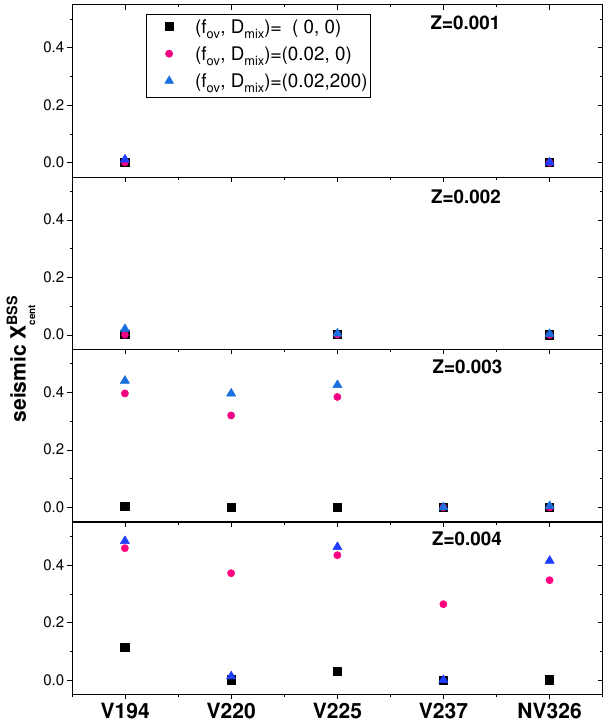}
	\caption{Central hydrogen abundance, $X_{\rm cent}$, of the collisional seismic models of the five SX Phe stars from $\omega$ Cen, for four values of metallicity, $Z$. Different symbols indicate the adopted mixing prescriptions, $(f_{\rm ov},D_{\rm mix})$.}
	\label{figC1}
\end{figure}

\end{appendix}

\end{document}